# Transparent Photothermal Metasurfaces Amplifying Superhydrophobicity by Absorbing Sunlight

*Efstratios Mitridis, Henry Lambley, Sven Tröber, Thomas M. Schutzius\*, Dimos Poulikakos\**

Laboratory of Thermodynamics in Emerging Technologies, Department of Mechanical and Process Engineering, ETH Zurich, Sonneggstrasse 3, CH-8092 Zurich, Switzerland

\* To whom correspondence should be addressed.

Prof. Dimos Poulikakos
ETH Zurich
Laboratory of Thermodynamics in Emerging Technologies
Sonneggstrasse 3, ML J 36
CH-8092 Zürich
SWITZERLAND
Phone: +41 44 632 27 38
Fax: +41 44 632 11 76
dpoulikakos@ethz.ch

Prof. Thomas Schutzius
Current address:
ETH Zurich
Laboratory for Multiphase Thermofluidics and Surface Nanoengineering
Sonneggstrasse 3, ML J 27.2
CH-8092 Zürich
SWITZERLAND
Phone: +41 44 632 46 04
thomschu@ethz.ch

**Abstract**

Imparting and maintaining surface superhydrophobicity is receiving significant research attention over the last several years, driven by a broad range of important applications and enabled by advancements in materials and surface nanoengineering. Researchers have investigated the effect of temperature on droplet-surface interactions, which poses additional challenges when liquid nucleation manifests itself, due to ensuing condensation into the surface texture that compromises its anti-wetting behavior. Maintaining surface transparency at the same time poses an additional and significant challenge. Often, the solutions proposed are limited by working temperatures or are detrimental to visibility through the surface. Here we introduce a scalable method employing plasmonic photothermal metasurface composites, able to harvest sunlight and naturally heat the surface, sustaining water repellency and transparency under challenging environmental conditions where condensation and fogging would otherwise be strongly promoted. We demonstrate that these surfaces, when illuminated by sunlight, can prevent impalement of impacting water droplets, even when the droplet to surface temperature difference is 50°C, by suppressing condensate formation within the texture, maintaining transparency. We also show how the same transparent metasurface coating could be combined and work collaboratively with hierarchical micro- and nanorough textures, resulting in simultaneous superior pressure-driven impalement resistance and avoidance of water nucleation and related possible frosting in supercooled conditions. Our work can find a host of applications as a sustainable solution against impacting water on surfaces such as windows, eyewear, and optical components.

Inspired by natural examples (lotus leaf,[1] butterfly wings,[2] water strider legs[3]), superhydrophobic surfaces have received great attention in recent years for their unique self-cleaning,[4] anti-fouling,[5] and anti-icing[6] properties. Their extreme water repellency, especially with respect to impacting droplets, can be achieved by introducing open and closed microstructures,[7–9] hierarchical surface roughness,[10–14] low-surface-energy materials,[15–18] and substrate flexibility.[16,19,20] Such surface and substrate properties act to stabilize an intervening, lubricating air layer that is responsible for high droplet mobility (so-called Cassie-Baxter wetting state[21]) and prevent a transition to the sticky Wenzel wetting state (*i.e.* impalement).[22] For real-world applications, the ability to repel impacting droplets is critical with numerous studies having investigated it under a range of environmental conditions and droplet temperatures ranging from supercooled droplets[20] and surfaces[7,23] to ambient[24,25] and hot droplets.[26] Typically, impalement is prevented by ensuring that the anti-wetting (capillary) pressure is greater than the wetting (droplet) pressure.[27] However, a host of additional mechanisms responsible for the loss of superhydrophobicity emerge when departing from ambient conditions including condensation within the microtexture when exposed to hot vapor[26,28–30] and increased viscosity and freezing in the case of supercooled droplets.[7,20,31–33] Condensation resistant superhydrophobic surfaces have been demonstrated using nanotexturing through droplet coalescence,[34,35] scaling the texture to prevent filling within the droplet contact time[36,37] and minimizing the adhesion force of filled cells.[26] Yet, all of these approaches fail to prevent the initial nucleation of condensate embryos, limiting their working envelope, especially for high supersaturation conditions. This is a critical issue when imparting nanotexture to a surface, which is routinely done to enhance impalement resistance by increasing the antiwetting capillary pressure. More specifically, when using nanotexture, due to its significantly smaller dimensions compared to microtexture, if a droplet does nucleate within the texture, it fills the roughness very quickly and displaces the lubricating air layer. Consequently, in a supersaturated environment, despite the fact that the nanotexture increases the pressure-based impalement resistance, it can actually reduce the repellency of the surface[38] due to its vulnerability to water nucleation. Moreover, providing multi-functionality by maintaining optical transparency,[39,40] an important property in many practical applications, is inherently counteractive to imparting superhydrophobicity, since surface coating roughness inherently obscures light by causing scattering.[41]

To address these challenging intertwined problems of condensation-enhanced impalement and the limitations of surface energy and texturing on resisting impalement, while maintaining transparency, we propose an additive approach exploiting unexplored aspects of the involved physics. Inspired by other innovative studies using sunlight for path-breaking applications,[42–46] including catalysis,[47] desalination,[48,49] materials synthesis,[50] and ice/fog repellency,[51,52] we present rationally designed, transparent yet solar light absorbing, superhydrophobic plasmonic metasurface composites.[51,52] Guided by mass diffusion, heat transfer, and nucleation theories, we investigate the effects of microtexture and nanotexture, substrate subcooling, droplet impact speed, and light intensity on their photothermal operating range and performance. We explore the repellency of water droplets and the mechanisms of texture cavity filling due to condensation across a broad range of temperatures—all while maintaining optical transparency, critical to many functional water repellent surfaces exposed to natural light. We also demonstrate how our coating works in symbiosis with nanotexture to offset inherent undesirable nucleation, which causes failure under moderate supersaturation conditions.

**Results and Discussion**

To study the ability of metasurfaces to amplify superhydrophobicity, we fabricated transparent polyurethane acrylate (PUA) micropillared patterns using soft lithography on glass substrates. On top of these, we deposited a well-adhering, ultra-thin metasurface coating with sputter deposition.[51] This coating, relying on surface plasmon effects, significantly enhances broadband light absorption and heating, achieved by nanoengineering of gold nanoparticle inclusions in a titanium dioxide matrix (dielectric); the mean levels of transparency and absorption are controlled by coating thickness.[51,52] The broadband absorption property is achieved by tuning the size distribution and concentration of the inclusions in titanium dioxide. For concentrations close to the percolation limit (*i.e.* adjacent nanoparticles forming a continuous electrically conductive path), electronic conductivity increases dramatically, leading to very effective photothermal light absorption. This is important for harvesting the infrared fraction of sunlight, which makes up more than 50% of its total power, and in contrast to tinted window films and automobile glass, which work by—instead—reflecting infrared.[53] In a subsequent step, we deposited an ultra-thin conformal low surface energy fluoropolymer coating with initiated chemical vapor deposition (iCVD), resulting in a superhydrophobic plasmonic metasurface composite. Both the metasurface and fluoropolymer coatings do not substantially alter the underlying geometrical features of our micropillared surfaces. (See Methods, Supporting Information, section "Topography of the control", and Figure S1 for fabrication steps of all types of surfaces used in this study and topography of a superhydrophobic control surface.)

Figure 1a shows a scanning electron micrograph of the plasmonic composite surfaces used in this study. We observe that both the metasurface and fluoropolymer coatings cover the micropillared substrate uniformly, making them almost indistinguishable from the bare pillars. Especially for the metasurface coating, we used low-pressure and high-power sputtering to promote deposition directionality (see Methods for deposition conditions), which ensures coverage of the bottom of the texture, and also to enhance conformality through metal ion resputtering on the sidewalls of the micropillars.[54] Some submicron features on the surface are due to the soft lithography replication process, but do not negatively impact its performance. (For more information on the superhydrophobic control surface, see Supporting Information, section "Topography of the control", and Figure S1.) We measured the apparent static, advancing and receding water contact angle values on the surface to be $\theta^* = 154° \pm 2°$, $\theta_a^* = 160° \pm 2°$ and $\theta_r^* = 141° \pm 3°$, respectively. For reference, we measured the respective contact angles of a glass substrate coated with a fluoropolymer to be $\theta^* = 122° \pm 2°$, $\theta_a^* = 125° \pm 1°$ and $\theta_r^* = 119° \pm 1°$. Figure 1b shows a schematic of the different layers of our samples: glass substrate, PUA micropillars, metasurface coating, and fluoropolymer coating. The micropillar array has pillars with diameter, $d$, spacing, $s$, and height, $h$: $[d, s, h] = [2.5, 5.0, 3.3]$ μm. Figure 1c shows a cross-sectional micrograph of a representative metasurface coating with exaggerated thickness (for imaging purposes), revealing gold nanoparticles (bright spots) embedded in a titania matrix. The high optical transparency of our plasmonic composite is evident in the pictures of Figure 1d, where a printed logo was placed directly underneath it, also for the case of a control surface (lacking the metasurface coating), and illuminated from behind.

Figure 1e shows a plot of the transparency, $\mathcal{T}$, reflectivity, $\mathcal{R}$, and absorption, $\mathcal{A}$, vs wavelength (visible and near infrared; 400–2200 nm) for normal light incidence. The mean values of $\mathcal{T}$, $\mathcal{R}$, and $\mathcal{A}$ were 62%, 14% and 24%, respectively, indicating that one can balance transparency with absorption using transparent metasurface coatings ($\mathcal{T} = 51\%$ in the visible; not opaque). (For more information on the optical properties of the control, see Supporting Information, section "Transparency, reflectivity and absorption of superhydrophobic control surfaces", and Figure S2.) Figure 1f shows a schematic of the experimental setup used to quantify the effect of illumination on heating of the surface. It consists of a light source (halogen lamp or solar simulator, for low or high power density, respectively; see Methods); an infrared (IR) camera and a pair of thermocouples that measure the sample, $T_s$, and ambient, $T_\infty$, temperatures. The normalized irradiance spectra of both light sources are shown in the inset. We note that the plasmonic composites absorb at wavelengths between 400 and 2500 nm, which is practically the energy range of the entire solar spectrum. Figure 1g shows plots of $T_s - T_\infty$ vs time, $t$, for surfaces illuminated with a solar simulator (power densities: $P$ = 1 kW m$^{-2}$ and 3.5 kW m$^{-2}$; used later in this work). At $t$ = 300 s, we see that, for the plasmonic composite, $T_s - T_\infty$ = 4°C and 16°C for $P$ = 1 kW m$^{-2}$ and 3.5 kW m$^{-2}$, respectively, while, for the same $P$, the $T_s - T_\infty$ of control surfaces is at least 3 times lower, demonstrating the significant role of the plasmonically captured sunlight in heating through the photothermal effect. We found that our hydrophobized metasurface adhered very well to substrates after standard tape peel and abrasion tests, and that it still retained its heating property even after partial removal due to scraping (see Supporting Information, section "Mechanical durability of the metasurface coating", and Figure S3). In the following, to avoid unwanted condensation from humidity in the environment, we heat the droplets instead of cooling the substrates to achieve supersaturation. (See also Supporting Information, section "Inhibiting fog with sunlight in plasmonic composite surfaces", and Figure S4.)

Figure 2a shows an image sequence of a water droplet with an initial radius, $R_0 \approx 1.6$ mm, density, $\rho$, surface tension, $\gamma$, and temperature, $T_w = 70°C$, impacting onto our superhydrophobic plasmonic composite with a velocity, $v \approx 0.8$ m s$^{-1}$, without illumination ($T_\infty = 21°C$). We define the Weber number, which describes the relative importance of inertia and surface tension during droplet impact, as $We = 2\rho v^2 R_0 / \gamma$. For these experiments, $We$ was kept constant at 26. The chosen $T_w - T_\infty$ means that there is a high supersaturation above the surface and a high probability of condensation under normal circumstances. Figure 2b depicts a droplet impacting onto the same surface under identical conditions except that now the surface is illuminated (solar simulator; power density, $P$ = 1 kW m$^{-2}$). In Figure 2a, we see that the droplet leaves a remnant on the surface (characteristic of impalement), while in Figure 2b the entire droplet rebounds from the surface (see Supporting Information Video S1). Figure 2c plots the probability of droplet impalement on the plasmonic composite, $\Phi$, vs $T_w$, for $P = 0$ and 1 kW m$^{-2}$. For $T_w = 40°C$, we note the onset of impalement for $P = 0$ kW m$^{-2}$ while for $P = 1$ kW m$^{-2}$ we observe no impalement. Therefore, we see that sunlight alone is capable of boosting the existing repellency of a superhydrophobic surface. We attribute impalement to water condensation within the microtexture due to the evaporation of the warmer droplets whose vapor displaces the intervening air layer, as discussed in previous works.[26,29] (See also Supporting Information, section "Enhanced impalement resistance with concentrated light" and Figure S5 for the effect of concentrated illumination on impalement resistance.)

To further understand and quantify this repellency boosting mechanism, we study next the wetting and impalement behavior of sessile warm droplets as a function of incident light power density, $P$. Figure 3a, Figure 3b, and Figure 3c show image sequences of such droplets with the same initial temperature ($T_w = 55°C$) for $P = 0$, 1, and 2 kW m$^{-2}$ (halogen lamp), respectively. The different values of $P$, 0, 1, and 2 kW m$^{-2}$, result in different surface temperatures, $T_s$, before droplet placement, of 22°C, 24°C, and 28°C, respectively. We define $t_f$ as the duration between placement of a warm droplet onto the surface and the onset of contact line motion. In all previous image sequences, we observe motion for $t = t_f$, accompanied by a decrease in the apparent contact angle and an increased droplet-surface contact area; we presume that this is equivalent to impalement. A clear trend of increasing $t_f$ with increasing $P$ exists. To better investigate this behavior, we can theoretically calculate the approximate timescale needed for condensate, which is generated by the surface being colder than the droplet, to grow and fill the texture. Assuming that this process is diffusion limited, the theoretical filling timescale for constant $T_s$ and $T_w$ can be expressed as[26]

$$\tau_f = \rho h^2 / D_v \Delta C \qquad (1)$$

where $D_v$ is the water vapor diffusion coefficient in air and $\Delta C$ is the water vapor concentration difference between the droplet and the bottom of the texture. $\Delta C$, in turn, is a function of the local water vapor saturation pressure difference across a cavity, $\Delta p_{v,sat}$, which is a function of both the droplet and substrate temperatures. It is noted that $\tau_f$ strongly depends also on the pillar height, scaling as $h^2$. (For more details on $\tau_f$, see Supporting Information, section "Theoretical basis of impalement criterion".)

Figure 3d shows a plot of $t_f$ vs $\tau_f$ for three different values of $P$ (0, 1, and 2 kW m$^{-2}$). Lines of best fit are shown for these three cases. We varied the values of $\tau_f$ by varying $T_w$; these two quantities have an inverse correlation. Figure 3d provides a quantitative relationship between $t_f$ and $\tau_f$, from which we define $\alpha = t_f / \tau_f$ as an empirical correlation coefficient, and their dependence on $P$. In relative agreement with previous research on hot droplets interacting with ambient substrates,[26] we found $\alpha|_{T_s=T_\infty=22°C} \approx 11$. However, for $T_s > T_\infty$, which occurs naturally due to the surface being illuminated, we observed that $\alpha$ increases significantly to $\alpha|_{T_s=28°C} \approx 26$ (see Supporting Information, sections "Effect of surface temperature on condensation nucleation on superhydrophobic control surfaces" and "Experimental setup, water temperature calibration and droplet cooling rate", Figure S6 and Figure S7). Figure 3e shows schematically the mechanism of droplet impalement due to condensation from a warm droplet over time inside the surface texture. For contact times, *i.e.* the duration of the touch of the droplet on the surface, $t_c \ll t_f = \alpha \tau_f$, the cavities are almost water-free and superhydrophobicity is maintained. As $t_c$ and $\alpha \tau_f$ become comparable, condensation occurs and the cavity begins to fill with water, while for even longer $t_c$, water completely fills the cavity and local superhydrophobicity is compromised.

To explore a potential coupling of pressure and condensation-based impalement mechanisms, we mapped the impalement probability, $\Phi$, of warm droplets impacting onto our plasmonic composites for a range of $We$. The image sequences in Figure 4a and Figure 4b depict droplets with $T_w > T_s$ impacting onto a surface at $We = 73$ for $P = 0$ and 3.5 kW m$^{-2}$, respectively. We used a concentrated illumination of 3.5 suns to better demonstrate any possible effects of light on condensation-based impalement. We see that even though the droplet is warmer in the case of the illuminated sample, and therefore more likely to impale, the value of $P$ used here is sufficient to suppress impalement and boost the naturally existing repellency of the surface. To isolate the pressure-based impalement mechanism, we performed droplet impact experiments in isothermal conditions ($T_w = T_s = T_\infty$) for $We = 73$, which revealed a significant impalement probability, $\Phi \approx 86\%$. (See Supporting Information, section "Effect of light on pressure-driven impalement in plasmonic composite surfaces" and Figure S8. See also section "Comparison of droplet retraction dynamics" and Figure S9 for the relationship between droplet recoil dynamics and outcome of an experiment, impalement *vs* rebound.)

In order now to develop a theoretical predictor of dynamic droplet impact impalement events, either pressure or condensation-based, we additionally define the theoretical droplet oscillation time as[20] $\tau = (\pi/4)\sqrt{\rho R_0^3 / \gamma}$, which is a measure of droplet-substrate contact time during impact (see also Supporting Information, section "Theoretical basis of impalement criterion"). A comparison between the droplet contact and cavity filling theoretical timescales, where the latter has been corrected by the previously calculated correlation coefficient ($\alpha$), $\tau/\alpha\tau_f$, constitutes our predictive impalement criterion. Figure 4c shows a plot of $\Phi$ *vs* $\tau/\alpha\tau_f$, for warm droplet impact events at $We = 26$ and 73, for light off and light on cases, $P = 0$ and 1 or 3.5 kW m$^{-2}$, respectively. When $T_w$ is increased, $\tau/\alpha\tau_f$ also increases. Were the model proposed to capture all the physics involved, impalement transition would be expected to occur for $\tau/\alpha\tau_f \approx 1$, with full impalement ($\Phi = 100\%$) predicted at values $\geq O(1)$. The lines for different experimental conditions should collapse into one. In practice, we observe this behavior in the onset of impalement transition for both light off and light on cases at low $We$, for $\tau/\alpha\tau_f \in [10^{-2}, 1)$; the maximum droplet temperature we could repeatably achieve experimentally limits us to values <1. For the high $We$ cases, however, the onset is at $\tau/\alpha\tau_f \approx 10^{-2} < 1$, with concentrated illumination mildly decreasing $\Phi$. This behavior can be explained through the schematics in Figure 4d and Figure 4e. At low $We$, the condensation-driven mechanism is dominant and impalement can be fully predicted by $\tau/\alpha\tau_f$. However, at higher $We$, a high wetting pressure results in meniscus penetration, to a depth, $\delta$, into the texture during impact, effectively reducing the cavity height needed to be filled with water for impalement to occur. This reduces $\tau_f$, which—in this case—is $\sim (h-\delta)^2$, resulting in an underestimated $\tau/\alpha\tau_f$. We term this mechanism combined, as there are condensation and pressure components acting simultaneously. It would be possible to use the same impalement criterion for the combined mechanism by replacing $h$ with $h-\delta$. For example, if $\delta = 0.8 \cdot h$, $\tau/\alpha\tau_f$ increases twenty-five-fold; this would shift impalement transition towards unity. Regardless, for the scope of this study, our criterion holds well for order of magnitude analyses involving cavity filling. (See also Supporting Information, section "Mapping droplet impact behavior on plasmonic composite surfaces", and Figure S10 for a comprehensive dynamic droplet repellency performance map of plasmonic composites.)

We will now explore strategies to fundamentally address the pressure-based impalement mechanism. Figure 5a and Figure 5b show image sequences of ambient droplets with $T_w = T_\infty = 21°C$ impacting onto our control surface (no metasurface coating) for $We = 26$ and $We = 73$, respectively; the surfaces are not illuminated ($P = 0$ kW m$^{-2}$). We observe droplet rebound in the low $We$ case, while, in contrast, at $We = 73$, a daughter droplet remains on the surface owing to the pressure driven impalement mechanism. To address this, we improved the superhydrophobicity of the single-tier plasmonic composite by spray coating a dispersion of silanized (hydrophobic) silica nanoparticles with a polymeric binder for enhanced durability. Adding a second tier of roughness does not compromise surface transparency; see inset logo picture in Figure 5c. Figure 5c and Figure 5d depict water droplets of temperature, $T_w = 21°C$ and $32°C$, respectively, impacting onto our hierarchical plasmonic composite for $We = 95$, a more vigorous test. We notice the effect of nanoroughness in the increased impalement resistance, evident by the full repellency of the ambient water droplet ($n = 7$ experiments) at high $We = 95$, Figure 5c. However, raising $T_w$ to $32°C$ leads to 100% impalement probability ($n = 7$), Figure 5d. Finally, Figure 5e shows how effectively sunlight ($P = 1$ kW m$^{-2}$), for the same $We$ and $T_w$ as in Figure 5d, helps mitigate cavity filling and impalement by drastically reducing $\Phi$ to 28%.

In order to explain the previous—unexpected—impalement events and also the effect of illumination, we investigated the nanocavity filling dynamics. From an extended view of the inset nanoroughness micrograph in Figure 5c-e, and assuming for simplicity cylindrical nanopores with an aspect ratio of unity, we measured the average nanopore diameter and subsequently depth to be $\approx 50$ nm. To determine the dominant water vapor mass transfer mechanism at these small length scales, we define the Knudsen number, $Kn = \lambda_f / d_r$, which compares $d_r$ to the molecular mean free path in saturated moist air, $\lambda_f$. For saturated ambient conditions,[55] $Kn \approx 1.5$, rendering a continuum mechanics approach inapplicable, since for $Kn > 0.5$, the dominant mechanism is free molecular flow.[56] The correct theory for describing such flows is based on effusion,[57] and with this the lower bound of the nanopore filling time, $\tau_{f,e}$, is very short ($\sim 10^{-4}$ s). (See Supporting Information, section "Effusion in a nanopore" for the calculation method of the nanopore filling time due to effusion.) For $T_w = 32°C$ and $T_\infty = 21°C$, we calculate $\tau/\tau_{f,e} \approx 10^3 \gg 1$, indicative of impalement, considering effusion alone. For nanopore filling to be initiated, we must verify that condensation nucleation occurs at a significant rate. Based on classical nucleation theory, Figure 5f plots $P$ vs $T_w - T_\infty$ vs $N$, where $N$ ($= \tau A J$, where $J$ is the heterogeneous condensation nucleation rate [nuclei per area per time] and $A$ is the surface area of a single cylindrical nanopore) is defined as the number of water nuclei per nanopore, formed during droplet impact. $T_w$ was kept constant at 32°C and $\Delta T = T_w - T_\infty$ was varied by changing the temperature of the colder environment, $T_\infty < T_w$. (See also Supporting Information, section "Heterogeneous condensation nucleation".) A local static water contact angle, $\theta = 120°$, corresponding to minimum wetting in the nanoscale, due to the hydrophobic nanoparticles, was considered. We calculated the radius of curvature of the nanopits as 4 nm using the Brunauer–Emmett–Teller theory for the silica nanoparticle powder and assuming spherical and monodisperse nanoparticles. Furthermore, a linear relationship between $P$ and surface temperature applies. Defining $N = 1$ [nuclei per nanopore per contact time] as probable for condensation nucleation based impalement, the non-illuminated surface can only sustain a non-wetting state until $\Delta T \approx 8°C$, which explains the impalement events we observed (Figure 5d). For this $\Delta T$, nanopore filling is nucleation limited—in contrast to the single-tier geometry. Solar illumination of one sun boosts $\Delta T$ by 4°C and prevents nucleation in the first place, which, in turn, prevents impalement (Figure 5e).

The advantages of our plasmonic composite coating extend beyond ambient conditions into supercooled environments, where freezing is also probable to occur if condensation is not controlled. For $T_w = 0°C$, we compute a minimum working $T_\infty$ of -7°C without and -11°C with solar illumination. This 4°C shift is very important, considering the exponential dependence of condensation nucleation rate, $J$, on surface temperature. In fact, heating by just 1°C can reduce nucleation rates by up to several orders of magnitude for both condensation and freezing processes,[51,52,58] emphasizing how sunlight can be effectively and passively employed to prevent impalement on surfaces that typically forgo condensation resilience in pursuit of superior pressure-based impalement resistance.

**Conclusions**

We demonstrated a facile and passive method for achieving superior water repellency while maintaining surface transparency in supersaturated environments, relying on the collaborative effect of superhydrophobicity and passive heating through plasmonic exploitation of the photothermal effect stimulated by sunlight. Harvesting solar power with ultrathin plasmonic metasurface coatings without being detrimental to transparency, we demonstrated that typical microstructures can sustain superhydrophobicity and avoid droplet impalement at substrate temperatures much lower than the droplet temperature by fully preventing or retarding cavity filling in the diffusion-limited regime due to light absorption heating. We then discovered that adding a second tier of roughness to our micropillars, intended to boost the superhydrophobicity of the surface against high Weber number (impact velocity) incoming warm droplets, counterintuitively caused the surface to be less repellent by shifting cavity filling to a nucleation-limited regime. We established that by using a metasurface coating, one could inhibit condensation nucleation and significantly expand the working envelope of such hierarchical superhydrophobic surfaces, especially for supercooled environments, where it would have the added benefit of reducing the likelihood of freezing. Our approach can be used as a stand-alone or additive method towards counteracting the negative effects of warm(er) water vapor condensation on cold(er) surfaces and whenever a good degree of optical transparency is required. We believe that our work can make its advantages evident in a plethora of applications including glasses, optical components, windows, and windshields exposed to warmer humid conditions.

**Methods**

*Materials*

Microscope glass slides (76 by 26 mm, 1-mm thick, with cut edges and plain end) were obtained from VWR. Polydimethylsiloxane (PDMS) silicone elastomer and curing agent (Sylgard® 184, 10 to 1 ratio) were purchased from The Dow Chemical Company, and a UV-curable polyurethane acrylate (PUA) resin (MINS-311RM), composed of a functionalized pre-polymer, a photoinitiator and a UV-curable releasing agent, was obtained from Minuta Technology. Trichlorovinylsilane (TCVS, 97%), 1H,1H,2H,2H-perfluorodecyl acrylate (PFDA, 97%, with tert-butylcatechol inhibitor), tert-butyl peroxide (TBPO, 98%) and ethanol (96%, anhydrous) were sourced from Sigma-Aldrich. Hydrophobic fumed silica (HFS) nanoparticles (Aerosil® R 812) were purchased from Evonik Industries and Capstone® ST-100HS aqueous fluoropolymer dispersion from DuPont.

*Surface preparation*

*a. Plasmonic composite surfaces*. We fabricated a specially designed chrome hard mask consisting of 20 by 20 mm squares filled with circular dots with a target diameter of 2.5 μm and spacing ranging from ≈5 μm to ≈10 μm and used it to pattern 4-in silicon wafers with standard photolithography. We used a positive photoresist (AZ 4512; MicroChemicals GmbH) and the Süss MA6 mask aligner in hard contact mode for the UV exposure, and subsequently etched the wafers (deep silicon etch; PlasmaPro 100 Estrelas, Oxford Instruments) to a depth of $3.2 \pm 0.1$ μm. We then used initiated chemical vapor deposition (iCVD; iLab Coating System, GVD Corporation) to coat them with a thin and ultra-low surface energy pPFDA (9.3 mN m$^{-1}$) [59] conformal hydrophobic polymer layer. Upon a 10-min oxygen plasma ashing (100 W power; PCCE, Diener) of a wafer, we placed it in a glass chamber and ran a room temperature TCVS pre-treatment silanization process with initial pressure of ~13 kPa for 20 min, which would later enable chemical bonding of the pPFDA to silicon. For the iCVD deposition, the reactants were PFDA (monomer) and TBPO (initiator), and the parameters were the following: deposition time of 15 min, process pressure of ~13 kPa, PFDA flow rate of 1.0 sccm, TBPO flow rate of 2.6 sccm, and substrate temperature of 40°C. A heated filament at 300°C enabled breakdown of the reactants. The next step consisted of transferring the wafer pattern onto a flexible PDMS mold by means of soft lithography, for the preparation of which we mixed PDMS with a curing agent at a 10:1 ratio, poured it onto the silicon wafer and degassed under vacuum to remove all air bubbles. Curing was done in a convection oven at 70°C for 2 h. The mold, consisting of microholes, was peeled off the wafer. We then transferred the pattern from the flexible mold onto our substrates. We took microscope glass slides and thoroughly cleaned them, consecutively, in sonicated acetone, isopropyl alcohol and water baths, followed by plasma ashing for 10 min. Subsequently, we placed fractions of the PDMS mold onto thin layers of PUA resin, which we previously deposited on the glass slides. Curing and thus hardening of the pattern (micropillars) took place in a vacuum UV exposure box (Gie-Tec GmbH) for 10 min, followed by peeling off the PDMS molds. In order to deposit our light-absorbing metasurface coating on top of our structures, we employed sputter deposition (CS 320 C; Von Ardenne) method.[51] We first deposited ~20 nm of silicon dioxide ($SiO_2$) by RF-bias and setting the following parameters: 600 W power, 0.2 Pa pressure, 5 sccm argon gas flow and deposition time of 42 s. SiO$_2$ promotes surface wettability and adhesion of the succeeding metasurface coating, also increasing its homogeneity and reducing the overall nanoroughness. On top of it, we ran deposition cycles with alternating titania (TiO$_2$) and gold (Au) targets, up to a total number of 8 layers. For each Au deposition burst, we operated the tool for 3 s at 50 W DC-bias and 0.6 Pa pressure, while each deposition of TiO$_2$ took 43 s, using a 600 W RF field at 0.6 Pa pressure. The metasurface was terminated with an extra TiO$_2$ layer, deposited for 72 s. Finally, the metasurfaces were coated with a hydrophobic pPFDA layer in the same way as previously described in the case of silicon. Out of the pool of samples with different geometrical characteristics, we selected the ones with the following dimensions: $d \approx 2.5$ μm, $s \approx 5.0$ μm and $h \approx 3.3$ μm. The wetting fraction, $\varphi = \pi d^2 / 4s^2$, was ~20%.

*b. Control surfaces.* We fabricated the superhydrophobic control surfaces in almost the same manner as the plasmonic composite surfaces, with the only difference of omitting sputter deposition, meaning that curing of the PUA micropillars was directly followed by iCVD deposition of a hydrophobic pPFDA coating.

*c. Sample for metasurface coating cross-section depiction.* To demonstrate the multilayer nature of our metasurface coating, we deposited the metasurface similar to the aforementioned procedure on a smooth silicon substrate, omitting the SiO$_2$ layer, since no dewetting issues were observed in this case. We kept all deposition parameters the same and adjusted the total number of layers to 44.

*d. Hierarchical plasmonic composite surfaces.* We fabricated our hierarchical surfaces in two sets of steps. For the first roughness tier (micropillars) we followed the procedure described in *a*. For the second tier (nanoroughness), we prepared a ~2 wt.% of Capstone® ST-100HS polymer dispersion in ethanol, in which we added a ~1.3 wt.% dispersion of HFS nanoparticles, which we then sonicated for 3 min using a probe sonicator. The mean specific surface area of the HFS powder is 260 m² g⁻¹. We regulated the pH of the total dispersion to values of ~4 to ensure good dispersibility of both the polymer and HFS. Finally, we sprayed the dispersion onto our micropillars from a distance of 15-20 cm using a Paasche airbrush at an air pressure of 3 bar. Solvent evaporation was ensured by drying the surfaces on a hot plate set at 50°C for $>10$ min.

*Characterization*

We characterized the topography of the control, plasmonic composite and hierarchical surfaces by means of stylus profilometry (Bruker Dektak XT) and scanning electron microscopy (SEM; Hitachi SU8230). For the SEM micrographs of micropillars (with or without the nanorough coating), we set the acceleration voltage at 1-2 kV and utilized the secondary electron and lower detectors to collect surface and topographic information. For the cross-sectional micrograph, we selected the aforementioned detectors and raised the voltage to 5 kV. We measured all apparent static, advancing and receding contact angles with an OCA 35 goniometer (DataPhysics), with the tilting method and for droplet volumes of 10 μL. These angles were, in the case of the hierarchical plasmonic composite, $\theta^* = 159° \pm 3°$, $\theta_a^* = 167° \pm 1°$ and $\theta_r^* = 150° \pm 3°$, respectively. We carried out transparency, reflectivity and absorption measurements on our surfaces with the help of a V-700 (Jasco) UV-VIS spectrometer with an integrating sphere.

*Experimental setup and protocols*

We conducted surface temperature calibrations using either a high-speed infrared camera (SC7500, FLIR), for surfaces illuminated with halogen light (Flexilux 600 longlife), with $P = 1$ kW m⁻² and $P = 2$ kW m⁻², or a resistance temperature detector in the case of sunlike illumination with a xenon arc source (66902, Newport), equipped with an AM1.5 air mass filter to render it a solar simulator, for $P = 1$ kW m⁻² and 3.5 kW m⁻². The diameter of the light spot was ~8 mm for halogen light and >2 cm for the solar simulator. With the latter, we were able to homogeneously illuminate the whole sample. We assumed thermal equilibrium at $t = 300$ s (light was switched on at $t = 0$ s) and each measurement was run three times. Light from the solar simulator was stabilized by waiting for at least 30 min prior to running any experiments. (See Figure 1f for a schematic of the experimental configuration during calibration and Figure 1g for the temperature increase caused by the solar simulator on control and plasmonic composite surfaces.) We performed the warm droplet experiments when the illuminated surfaces reached thermal equilibrium and waited ~1 min between subsequent droplet dispensations. (See Supporting Information, section "Experimental setup, water temperature calibration and droplet cooling rate" and Figure S7, for more details on the experimental setups used for calibrating the water temperature and running the warm droplet experiments.) The droplet or impact position on a sample was shifted after every experiment to ensure minimal systematic effects of surface defects on our experiments and statistics. The relative humidity of the room was in the 40%–55% range throughout the study.

**Supporting Information**

The Supporting Information is available online free of charge. The following sections are included:

Topography of the control; transparency, reflectivity and absorption of superhydrophobic control surfaces; mechanical durability of the metasurface coating; inhibiting fog with sunlight in plasmonic composite surfaces; enhanced impalement resistance with concentrated light; theoretical basis of impalement criterion; effect of surface temperature on condensation nucleation on superhydrophobic control surfaces; experimental setup, water temperature calibration and droplet cooling rate; effect of light on pressure-driven impalement in plasmonic composite surfaces; comparison of droplet retraction dynamics; mapping droplet impact behavior on plasmonic composite surfaces; effusion in a nanopore; heterogeneous condensation nucleation; light-driven repellency of dynamic warm droplets (video).

**Author contributions statement**

**Acknowledgements**

Partial support of the Swiss National Science Foundation under grant number 162565 and the European Research Council under Advanced Grant 669908 (INTICE) are acknowledged. We thank Ute Drechsler (IBM Rüschlikon) for assistance in surface fabrication. We also thank Hyunchul Park for advice on UV-curable polymers and fabrication protocols, and Karl Schlichting (LTNT, ETH Zurich) for insights into effusion theory.

**Additional information**

**Competing financial interests:** The authors declare no competing financial interests.

# Figures

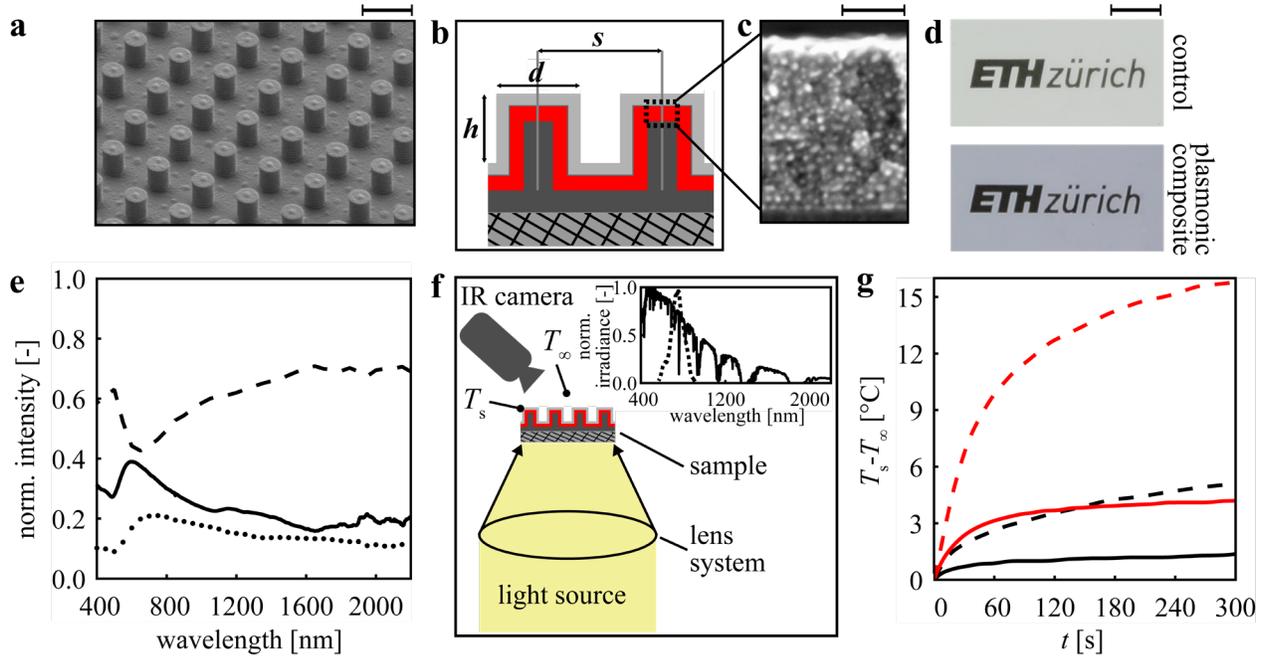

**Figure 1. Superhydrophobic plasmonic composite surface. (a)** Micrograph showing the surface from a tilted-view perspective. **(b)** Schematic of the surface cross-section, revealing its different layers. It consists of a hydrophobic fluoropolymer layer (top, light gray); plasmonic metasurface coating (red); micropillar array (dark gray); and glass substrate (gray, black hatch). **(c)** Micrograph showing the cross-section of a representative metasurface coating; the thickness here is exaggerated to facilitate imaging (the metasurface coating used throughout this study had a thickness of ~60 nm). The pillar diameter ($d$), height ($h$), and spacing ($s$) are shown in **b**. The sample in **a** has [$d$, $s$, $h$] = [2.5, 5.0, 3.3] μm. **(d)** The high transparency of the plasmonic composite surface in comparison to a control surface (lacks the metasurface coating) is evident in the readability of a printed logo placed directly underneath it. **(e)** Transparency (– – –), reflectivity (· · ·), and absorption (——) spectra of the surface shown in **a**. **(f)** Schematic of the experimental setup used to measure the temperature change of the samples relative to ambient upon illumination. The normalized irradiance spectra of the light sources we used in this study are shown as inset plots (solar simulator, ——; halogen lamp, - - -). **(g)** Surface temperature change relative to ambient, $T_s - T_\infty$, vs time, $t$, of control ($P$ = 1 kW m$^{-2}$, black curve; $P$ = 3.5 kW m$^{-2}$, black dashed curve) and plasmonic composite ($P$ = 1 kW m$^{-2}$, red curve; $P$ = 3.5 kW m$^{-2}$, red dashed curve) surfaces. The samples are illuminated when $t > 0$ s with a solar simulator. Each curve represents the mean values of $n$ = 3 experiments. The logos in **d** were reprinted with permission from ETH Zurich. Copyright 2020 ETH Zurich. Scale bars: **(a)**, 5 μm; **(c)**, 100 nm; **(d)**, 5 mm.

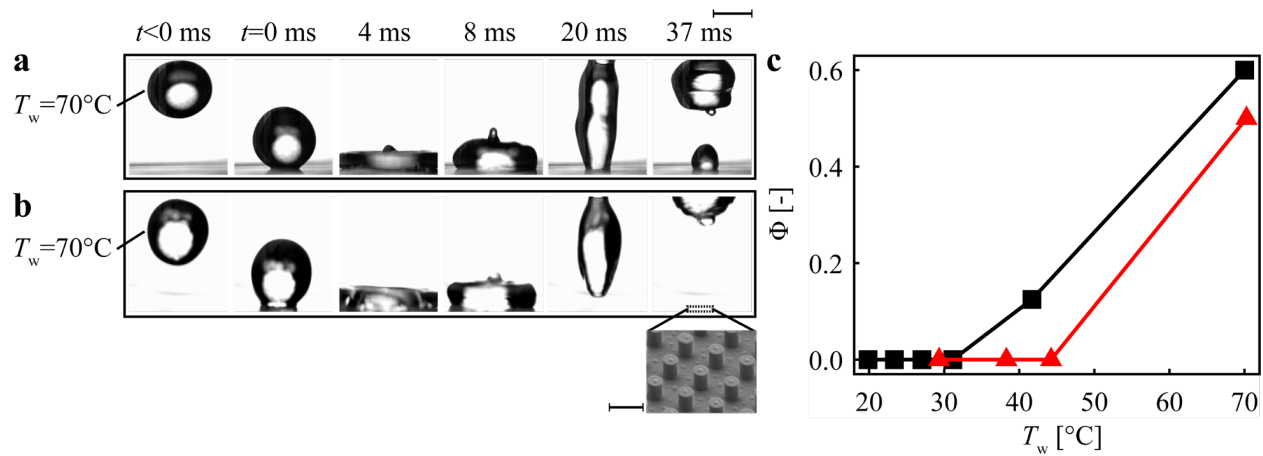

**Figure 2. Enhancing superhydrophobicity with one sun**. Image sequences showing warm droplets impacting onto a superhydrophobic plasmonic composite surface **(a)** without and **(b)** under illumination (solar simulator; $P = 1$ kW m$^{-2}$). For all experiments here, $We = 26$ and $T_\infty = 21\,°C$. A micrograph of the surface is shown as an inset image. **(c)** Probability of droplet impalement, $\Phi$, vs water droplet temperature, $T_w$, for no illumination (black squares) and under illumination (red triangles) conditions. Each data point represents $n \geq 7$ experiments. Scale bars: **(a)-(b)**, 2 mm; inset in **(b)**, 5 µm.

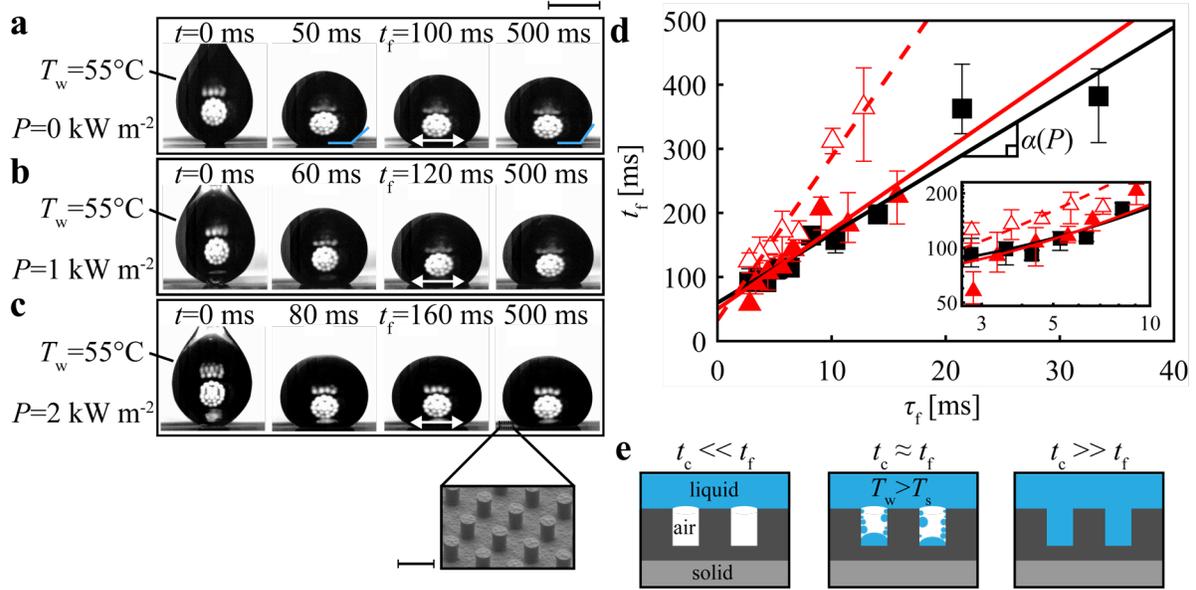

**Figure 3. Interplay between light coupling, plasmonic composite, and warm droplet repellency.** **(a)-(c)** Image sequences showing the wetting behavior of warm water droplets on a superhydrophobic plasmonic composite surface for **a**, $P = 0$ kW m$^{-2}$, **b**, $P = 1$ kW m$^{-2}$, and **c**, $P = 2$ kW m$^{-2}$ ($T_\infty = 22°C$). We used a halogen lamp to illuminate the samples. **(d)** Droplet impalement time, $t_f$, vs theoretical cavity filling time, $\tau_f$, for $P = 0$ kW m$^{-2}$ (black squares and line), $P = 1$ kW m$^{-2}$ (red filled triangles and red line), and $P = 2$ kW m$^{-2}$ (red open triangles and dashed red line). Each data point represents $n = 3$ experiments, and the lower and upper error values represent the minimum and maximum measured values, respectively. The three lines of best fit have slopes of $\alpha \approx 11$ (for confidence $C = 95\%$, $\alpha = [8,13]$), 12 ($C = 95\%$, $\alpha = [8,17]$), and 26 ($C = 95\%$, $\alpha = [18,33]$) corresponding to $P = 0$, 1, and 2 kW m$^{-2}$, respectively. The magnified inset, plotted in logarithmic axes, clarifies the behavior of the warmest droplets. **(e)** Schematic showing the mechanism of condensation impalement that occurs when a warm droplet is placed on the surface. Scale bars: **(a)-(c)**, 2mm; inset in **(c)**, 5 µm.
29

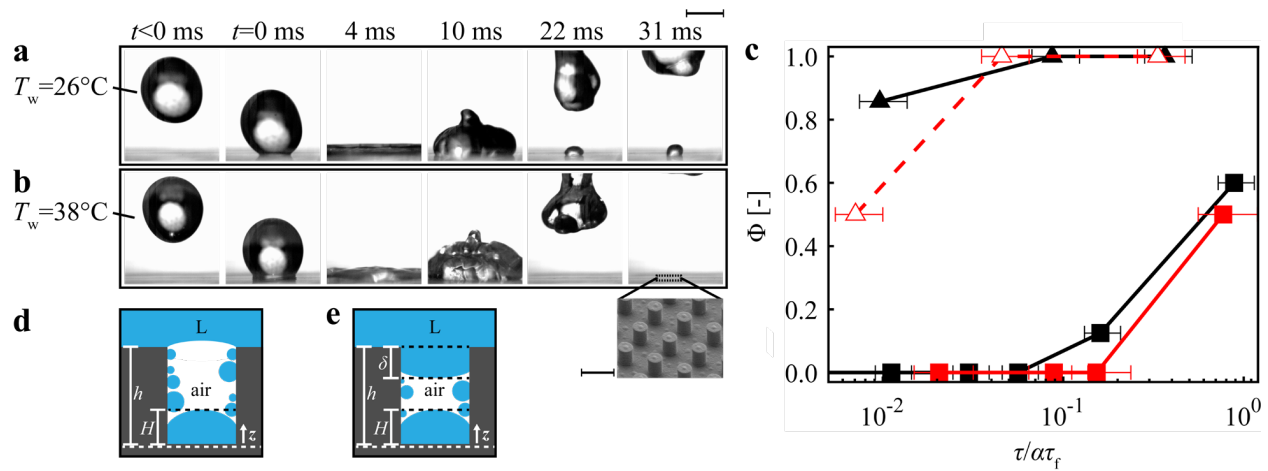

**Figure 4. Mechanisms of light-enhanced superhydrophobicity. (a)-(b)** Image sequences of warm water droplets impacting onto a superhydrophobic plasmonic composite surface ($T_\infty = 21°C$) for **a**, $We = 73$, $P = 0$ kW m$^{-2}$, and **b**, $We = 73$, $P = 3.5$ kW m$^{-2}$. A solar simulator was used for illumination. **(c)** Impalement probability, $\Phi$, vs ratio of droplet oscillation time to condensate filling time, $\tau/\alpha\tau_f$, for $We = 26$ and $P = 0$ kW m$^{-2}$ (black squares); $We = 26$ and $P = 1$ kW m$^{-2}$ (red filled squares and red line); $We = 73$ and $P = 0$ kW m$^{-2}$ (black triangles); and $We = 73$ and $P = 3.5$ kW m$^{-2}$ (red open triangles and dashed red line). The error bars are based on the 95% confidence level in the values of $\alpha$, which we previously calculated. Each data point represents $n \geq 7$ experiments. **(d)-(e)** Schematics of impalement mechanisms; **d**, condensation-driven, and **e**, condensation and pressure-driven (combined). Scale bars: **(a)-(b)**, 2 mm; inset in **(b)**, 5 μm.

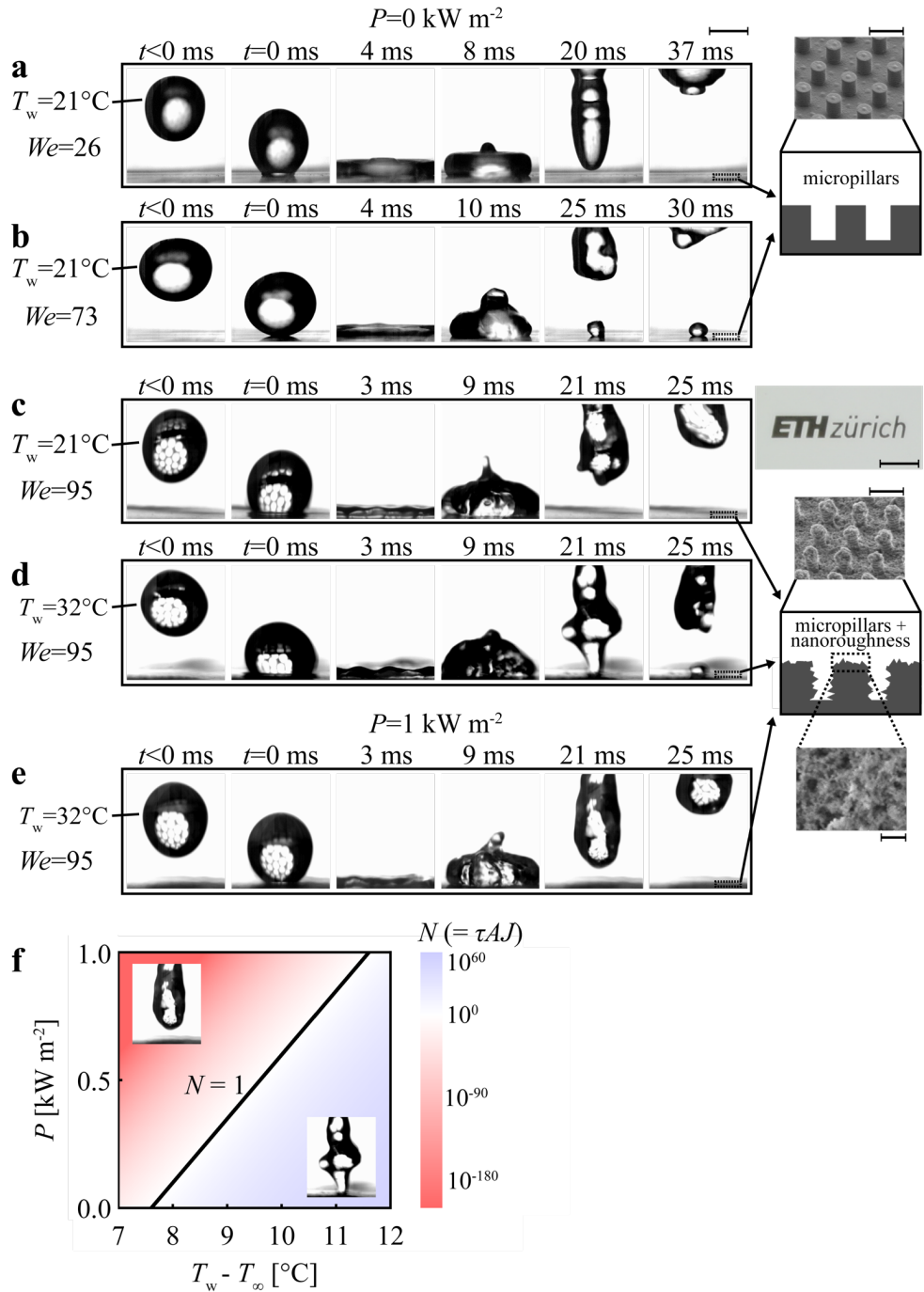

**Figure 5. Repellency boost with added nanoroughness and sunlight. (a)-(b)** Image sequences showing ambient water droplets ($T_\infty = 21°C$) impacting onto our control surface for **a**, $We = 26$, and **b**, $We = 73$. **(c)-(d)** Image sequences of water droplets impacting onto a hierarchical plasmonic composite for **c**, $We = 95$, $T_w = T_\infty = 21°C$, and **d**, $We = 95$, $T_w = 32°C$. In all cases, **a-d**, the surfaces were not illuminated ($P = 0$ kW m$^{-2}$). **(e)** Image sequence of a warm water droplet ($T_w = 32°C$) impacting onto the hierarchical surface for $We = 95$. Here the surface was illuminated ($P = 1$ kW m$^{-2}$). The inset in **c** shows a picture of the hierarchical plasmonic composite with a printed logo placed underneath it. Two insets in **a-e** show schematic representations and a scanning electron micrograph for each roughness tier. **(f)** Plot of incident solar power density, $P$, vs $T_w - T_\infty$, vs the number of water nuclei per nanopore, $N$ ($= \tau A J$), during droplet impact on the hierarchical plasmonic composite, for a fixed $T_w = 32°C$ and a range of $T_\infty < T_w$. $N = 1$ is considered as probable for condensation nucleation. Cylindrical pores of depth and diameter of 50 nm were assumed. The logo in **c**, inset, was reprinted with permission from ETH Zurich. Copyright 2020 ETH Zurich. Scale bars: **(a)-(e)**, 2 mm; inset in **(a)-(b)**, 5 μm; inset picture in **(c)**, 5 mm; inset in **(c)-(e)**, 5 μm (micropillars), 200 nm (nanoroughness).

**TOC**

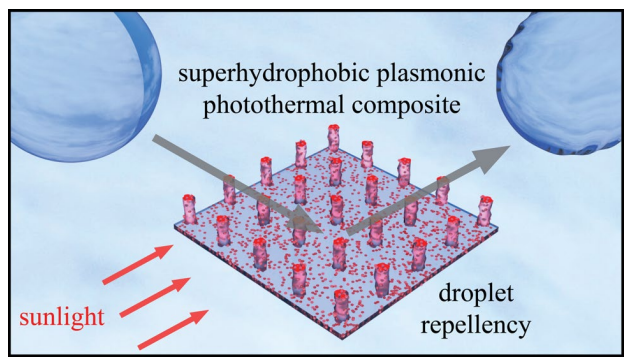

**Warm water repellency with sunlight and superhydrophobic plasmonic photothermal composite surfaces.**